\documentclass[12pt]{iopart}
\newfont{\mcal}{rsfs10 scaled 1200}
\usepackage{psfig,amssymb,graphics,epsfig}

\begin{document}
\hfill  CERN-2002-286 \\

\title[Particle production in time-dependent gravitational fields]{Particle 
       production in time-dependent gravitational fields: the 
       expanding mass shell}

\author{Sabine Hossenfelder,$^\dag$ Dominik J.~Schwarz,$^{\ddag *}$ 
        Walter Greiner$^\dag$} 

\address{${}^\dag$Institut f\"ur Theoretische Physik, J.W.Goethe-Universit\"at, 
         Robert-Mayer-Str.~8--10, 60054 Frankfurt am Main, Germany}

\address{${}^\ddag$Institut f\"ur Theoretische Physik, 
         Technische Universit\"at Wien, Wiedner Hauptstr.~8--10, 
         1040 Vienna, Austria}

\address{${}^*$Theory Divison, CERN, 1211 Geneva 23, Switzerland}

\begin{abstract}
We compute the production of particles from the gravitational field of an 
expanding mass shell. Contrary to the situation of Hawking radiation and 
the production of cosmological perturbations during cosmological inflation, 
the example of an expanding mass shell has no horizon and no singularity.
We apply the method of `ray-tracing', first introduced by Hawking, and 
calculate the energy spectrum of the produced particles. The result depends 
on three parameters: the expansion velocity of the mass shell, its radius, 
and its mass. Contrary to the situation of a collapsing mass shell, 
the energy spectrum is non-thermal. Invoking time reversal we reproduce 
Hawking's thermal spectrum in a certain limit. 
\end{abstract}

\section{Introduction}

Gradients in a gravitational field may cause the production of particles 
out of the vacuum; this is similar to the situation in quantum 
electrodynamics (QED). In the latter case a virtual pair of particles is 
ripped apart by the electric field. In contrast to QED, the gravitational 
`charge'---the mass---is always positive; in this case, the tidal forces 
are thus responsible for the separation of the particle pair. But there is 
also another difference. The energy that is needed for the production of 
particles in a charged vacuum is taken out of the field energy. Since 
the energy of the gravitational field cannot be localised, the energy gain 
must be explained otherwise. One possibility is the existence of a horizon. 
Another possibility is the time dependence of the gravitational field.

We compute the particle production in a time-dependent gravitational 
field induced by an expanding mass shell. This is done at a semiclassical 
level. The gravitational field, which acts through curvature of space-time, is 
treated as a classical background field in which a quantum field propagates. 
We consider a massless scalar field. So it is quantum field theory in curved 
space that we need to apply.

Perhaps the best known result in this field is the thermal radiation emitted 
by a black hole. This effect, discovered by Hawking \cite{Haw}, 
shows a deep connection between 
gravity, thermodynamics and quantum physics. Hawking's result was reproduced 
in several ways, but also other problems such as Unruh's accelerated observer 
\cite{Unruh} or the gravitational analogue to the Casimir effect 
\cite{Casm} were studied. 

In cosmological models, the production of particles plays a central role 
in explaining the origin of the structures observed in the Universe. During an
epoch of cosmological inflation, quantum fluctuations of space-time and 
matter are amplified by the accelerated expansion of the Universe to become
the seeds of large-scale structures \cite{infl}.

All mentioned examples involve either a horizon, a boundary, or a singularity.
To our knowledge, there exists no example in which particle production in 
an infinite space-time without horizon and without singularity has been
calculated. The situation of an expanding mass shell represents such an
example. We do not expect this work to have a direct astrophysical 
application, but the situation under study might be of interest when black 
holes evaporate. One could imagine that Hawking radiation is emitted 
from an evaporating black hole in a final burst, which we could then
describe as a thin shell, carrying a mass somewhat larger than the Planck mass.
If the emitted radiation consists of massive particles, this mass shell moves 
at a speed slightly below the speed of light and our calculation might 
be applicable. Since the expanding mass shell itself leads to gravitational 
particle production, this work might be useful for an estimate of backreaction
effects in the final stages of black hole evaporation. We will not study 
this aspect here, the purpose of this paper being just to calculate 
the energy spectrum of the particles produced by an expanding mass shell.   
 
There are three main procedures that are used to deal with quantum fields in 
curved space-time: Hawking's classical ray-tracing \cite{Haw},
renormalization of the energy-momentum tensor (e.g.~by point-splitting) 
\cite{Fulling,Wald}, particle production in the reference frame of a 
uniformly accelerated observer \cite{Unruh,GrQED},
from which we will choose the first one. A useful overview on 
computational matters can be found in \cite{BiDa,Novi}.

The procedure of ray-tracing in a semiclassical limit is appropriate if 
the energy carried by the quantum field is small with respect to the energy of 
the source of the gravitational field. Furthermore we have to stay on scales 
on which possible effects of quantum gravity do not occur.
As Wheeler showed in \cite{Pll} the characteristic length for this to 
happen is the Planck length $l_{\rm P}=
\sqrt{\frac{G\hbar}{c^3}}\approx 10^{-33}$ cm.

So we have to restrict our results to cases in which the effects of interest 
take place on scales $\gg l_{\rm P}$ and $\gg t_{\rm P}$. 
However, it seems that there is no reason to restrict the calculation to masses 
$M > m_{\rm P}$, as long as the mass shell starts to expand at a radius 
$\gg l_{\rm P}$. We set $G = c = \hbar = 1$ in the following.
 
This paper is organized as follows. First we will need to find an observable 
of particle density in curved space, which will be done in Section 2. After 
this we will introduce our model of the expanding mass shell and discuss its 
properties. The computation of the number density operator is presented in 
Section 4. The results are discussed in section 5, which contains plots of the 
spectral energy flux and total energy flux through a 2-dimensional surface. 
We conclude with a short summary. 

\section{Quantum fields in curved space-time}

Let us first recall some of the basics of quantum field theory in curved 
space-time \cite{BiDa,Wald,Fulling}. Physics in flat space-time is invariant
under Poincar\'{e} transformations. This singles out a set 
of (Fourier) modes and a vacuum state, which are invariant under Poincar\'{e} 
transformations. In curved space-time there is no global Poincar\'{e} symmetry.

In the following we consider a minimally coupled scalar field $\varphi$.
Its Fourier modes in flat space are attached to the energy $\omega$ by
\begin{equation} \label{Energ}
{\mathrm{i}} \partial_t u_{k}= \omega u_k \quad \mbox{with} \quad \omega>0\; . 
\end{equation}
This uniqueness is lost when the coordiates are chosen arbitrarily. 
In curved space-time we have no distinguished time coordinate, thus
no distinguished energy belonging to it, and no distinguished set of modes
that can be related to the notion of particles. In general we have to deal 
with an arbitrary set of modes $(u_i)$ and ask for its relation to a 
different set $(v_i)$. The quantum field $\hat{\varphi}$ can then be expanded 
in both sets:
\begin{equation} \label{Phiua}
\hat{\varphi} = \sum_i (u_i \hat{a_i}+u_i^*\hat{a_i}^{\dag})\; , \quad
\hat{\varphi} = \sum_j (v_j \hat{b_j}+v_j^*\hat{b_j}^{\dag})\; .
\end{equation}
The relations between these sets are given by the so-called
Bogoliubov transformations \cite{Bogo} and have the form:
\begin{equation} \label{Bogu}
v_j = \sum_i (\alpha_{ji}u_i+\beta_{ji}u_i^*) \;, \quad
u_i = \sum_j (\alpha_{ji}^*v_j-\beta_{ji}v_j^*) 
\end{equation}
with
\begin{equation} 
\alpha_{jk} = (v_j,u_k) \; , \quad  \beta_{jk}=-(v_j,u_k^*) \;,
\end{equation}
and the scalar product given by
\begin{equation} \label{skpr}
(u_i,v_j) =
-\mathrm{i}\int_{\Sigma}\left[u_i(\partial_{\mu} v_j^*) - 
                              (\partial_{\mu} u_i)v_j^*\right] 
           \sqrt{-g_\Sigma}\; \mathrm{d}\Sigma^{\mu}\;.
\end{equation}
Here $\Sigma$ is a space-like hypersurface with volume element 
$\mathrm{d}\Sigma$, $\mathrm{d}\Sigma^{\mu}=n^{\mu}\mathrm{d}\Sigma$, and 
$n^{\mu}$ is a normal vector of $\Sigma$ with norm $n^{\mu}n_{\mu}=1$.

The creation and annihilation operators (\ref{Phiua}) depend on the modes, 
and so does the vacuum state defined with respect to them. The vacuum state 
$|0_b\rangle$ belongs to $\hat{b}_i$ and is defined by 
$\hat{b}_i|0_b\rangle=0$. In general, $|0_b\rangle$ is not identical with
$|0_a\rangle$, for which we have $\hat{a}_i|0_a\rangle=0$. It follows that 
the vacuum expectation value of the number density operator 
$\hat{a}_i^{\dag} \hat{a}_i$ must not necessarily vanish when applied to 
the vacuum $|0_b\rangle$,  
\begin{equation} \label{Vakew}
\langle0_b|\hat{a}_i^{\dag} \hat{a}_i|0_b\rangle=\sum_j |\beta_{ji}|^2\;.
\end{equation}
This operator has a well defined meaning as the number (mode) density operator 
only if the modes used for the expansion can be put in a suitable relation to 
the known modes of Minkowski space-time to render an interpretation possible. 
This is possible if we are interested in the change of vacuum between two 
asymptotically flat regions of space-time.

If the eigenvalues are not discrete $(i \to \vec{k})$, we have to deal with 
an integral that may have divergences. These divergences caused by 
the limit from periodic boundary conditions with discrete spectrum to an 
infinite volume ${\bf{V}}=\langle 0|0\rangle$ can be cured by normalization. 
This may be accomplished by dividing through the same divergence:
\begin{equation} \label{stNprotau}
\langle0_b|\hat{a}_{\vec{k}}^{\dag} \hat{a}_{\vec{k}}|0_b\rangle_{ren} =
\frac{1}{\bf{V}}\langle0_b|\hat{a}_{\vec{k}}^{\dag} 
\hat{a}_{\vec{k}}|0_b\rangle=
\frac{1}{\bf{V}}\int |\beta_{\vec{k'}\vec{k}}|^2 \;{\mathrm{d}}^3 k' \;.
\end{equation}
If $\langle0_b|\hat{a}_{\vec{k}}^{\dag}\hat{a}_{\tilde{\vec{k}}}|0_b\rangle$ 
has the form $\delta(\vec{k}-\tilde{\vec{k}})f(\vec{k},\tilde{\vec{k}})$, 
one finds
\begin{equation} \label{Nprotau}
\langle0_b|{\hat{a}^{\dag}}_{\vec{k}} \hat{a}_{\vec{k}}|0_b\rangle_{ren}=
\frac{1}{\bf{V}}
\lim_{\vec{k} \to \tilde{\vec{k}}}\langle0_b|\hat{a}^{\dag}_{\vec{k}}
\hat{a}_{\tilde{\vec{k}}}|0_b\rangle=f(\vec{k})\; ,
\end{equation}
which will be useful later on. Now 
$\langle0_b|\hat{a}_{\vec{k}}^{\dag} \hat{a}_{\vec{k}}|0_b\rangle_{ren}$ 
is the well defined spectral number (mode) density.

\section{The expanding mass shell}

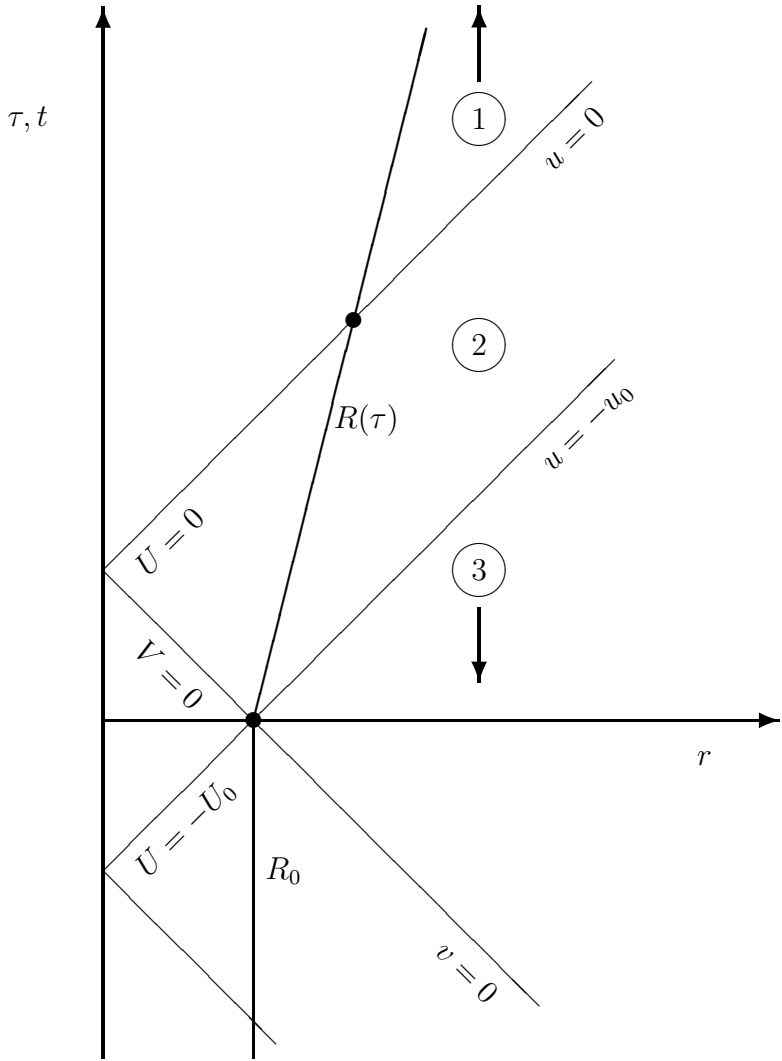
\begin{figure}
\begin{center}
\setlength{\unitlength}{1 cm}
\begin{picture}(12,15)


\put(1.0,0.0){\scalebox{2}{\vector(0,1){7.0}}}
\put(1.0,4.5){\scalebox{2}{\vector(1,0){4.5}}}

\thicklines
\put(3.0,0.0){\line(0,1){4.5}}
\put(3.0,4.5){\line(1,4){2.3}}

\thinlines
\put(1.0,6.5){\line(1,-1){5.8}}
\put(1.0,6.5){\line(1,1){6.5}}
\put(1.0,2.5){\line(1,-1){2.3}}
\put(1.0,2.5){\line(1,1){6.8}}

\put(6.0,6.0){\scalebox{2}{\vector(0,-1){0.5}}}
\put(6.0,13.0){\scalebox{2}{\vector(0,1){0.5}}}

\put(3.0,4.5){\circle*{0.2}}
\put(4.33,9.83){\circle*{0.2}}

\put(6.0,12.5){\circle{0.7}}
\put(6.0,9.5){\circle{0.7}}
\put(6.0,6.5){\circle{0.7}}

\put(6.8,11.8){\rotatebox{45}{$u=0$}}
\put(6.8,7.8){\rotatebox{45}{$u=-u_0$}}
\put(1.4,6.4){\rotatebox{45}{$U=0$}}
\put(1.4,5.4){\rotatebox{-45}{$V=0$}}
\put(1.4,2.4){\rotatebox{45}{$U=-U_0$}}
\put(5.4,1.4){\rotatebox{-45}{$v=0$}}

\put(6.0,12.5){\makebox(0,0){1}}
\put(6.0,9.5){\makebox(0,0){2}}
\put(6.0,6.5){\makebox(0,0){3}}

\put(4.5,8.5){\makebox(0,0){$R(\tau)$}}
\put(3.4,2.5){\makebox(0,0){$R_0$}}

\put(9,4){\makebox(0,0){$r$}}
\put(0.0,12.5){\makebox(0,0){$\tau,t$}}

\end{picture}
\end{center}
\caption{The expanding mass shell (thick line). The thin lines represent the 
rays at the border lines between the three cases of ray-tracing.} 
\label{scene}
\end{figure}

We consider a spherical shell of mass $M$ and radius $R$. Outside of the shell, 
space-time is described by the Schwarzschild metric; inside it space-time is 
flat. We assume that the thickness of the shell in negligible with respect to 
its radius (thin shell). Originally the shell's radius is constant, $R = R_0$.
The shell is `ignited' at some time $\tau = 0$ (as measured in the rest frame 
of the total mass inside the shell). Afterwards the shell expands as 
$R = R(\tau)$. The situation is illustrated in Fig.~\ref{scene}.

Our first task is to write down the metric of the space-time describing
an expanding mass shell. The surface of the shell is the same whether it is
measured from the inside or from the outside. We can thus choose the same 
radial coordinate $r$ on both sides of the shell and we obviously can 
choose the same angular coordinates $\theta$ and $\phi$. Outside the shell we 
denote the time coordinate by $t$, which is the time measured by an 
asymptotic observer, who is at rest with respect to the centre of mass of the 
shell. The relation between the time coordinates $t$ and $\tau$ remains to be 
fixed by the boundary conditions on the mass shell. We take null coordinates 
and ignore the $r^2 d\Omega^2$ part of the line element for the moment. Inside
the shell these coordinates are denoted by $(U,V)$, and by $(u,v)$ outside,
where the line element is given by
\begin{equation} 
\label{Meau}
\mathrm{d}s^2 = \gamma(r)\;\mathrm{d}u\;\mathrm{d}v\; , \quad 
\gamma(r) = 1-\frac{2M}{r}\; ,
\end{equation}
with
\begin{equation} 
\label{trafouv}
u + u_0 := t - \int^r_{R_0}\frac{\mathrm{d}r'}{\gamma(r')}\; , \quad 
v + v_0 := t + \int^r_{R_0}\frac{\mathrm{d}r'}{\gamma(r')}\; . 
\end{equation}
Inside we simply have 
\begin{equation}
\mathrm{d}s^2 = {\mathrm{d}}U\;{\mathrm{d}}V\; , 
\end{equation}
with
\begin{equation}
\label{trafoUV}
V + V_0 := \tau + r - R_0\; , \quad U + U_0 := \tau - r + R_0\; . 
\end{equation}
The outgoing light rays run on world lines with constant $u$ and $U$. 
We fix the origin of the null coordinates by the convention that the incoming 
ray $v=V=0$ goes through the point $(\tau = 0, R_0)$ and that $U=u=0$ denotes
the very same ray on its way out (see Fig.~\ref{scene}). Thus $v_0 = V_0 = 0$
and at $r=0$ we have $U + U_0 = V + 2R_0$. With our choice for $U$ we finally 
have $U_0 = 2R_0$.

We define the expansion velocity of the shell by
\begin{eqnarray}
\label{nu}
\nu := \frac{\mathrm{d}R}{\mathrm{d}\tau}\; ,
\end{eqnarray}
which is actually the coordinate velocity as seen from the inside of the 
shell. In principle $\nu$ is a function of time, but we will see below
that it is a good approximation to work with a constant expansion velocity.
In that case the world line of a point on the shell is given by
\begin{eqnarray} \label{RvonV}
R(\tau)=\left\{\begin{array}{lcr}
R_0 + \frac{\nu}{1+\nu}V =  R_0 + \frac{\nu}{1-\nu}(U+U_0)
&\mbox{when}& \tau >0\\
R_0 & \mbox{when} & \tau <0
\end{array}\right. \; .
\end{eqnarray}
Our result will finally depend on three parameters: $R_0$, $M$ and $\nu$. 

{}From the coordinate transformations (\ref{trafouv}) and (\ref{trafoUV}) and
from the definition of the expansion velocity (\ref{nu}) we find  
\begin{eqnarray}
\label{ugenU}
\left.\frac{\mathrm{d}u}{\mathrm{d}U}\right|_{r = R} &=&
\left\{\begin{array}{ccr}
       \left(\frac{\mathrm{d}t}{\mathrm{d}\tau} - \frac{\nu}{\gamma(R)}\right)
       \biggr/(1-\nu) 
         &\mbox{when}& U > -U_0 \\
       1/\sqrt{\gamma_0} &\mbox{when}& U < -U_0
       \end{array}\right.\; , \\
\label{ugenv}
\left.\frac{\mathrm{d}v}{\mathrm{d}V}\right|_{r = R} &=&
\left\{\begin{array}{ccr}
       \left(\frac{\mathrm{d}t}{\mathrm{d}\tau} + \frac{\nu}{\gamma(R)}\right)
       \biggr/(1+\nu)
         &\mbox{when}& V > 0 \\
       1/\sqrt{\gamma_0} &\mbox{when}& V < 0
       \end{array}\right.\; , 
\end{eqnarray}
where
\begin{eqnarray}
\gamma_0 := \gamma(R_0) = 1-\frac{2M}{R_0}\; .
\end{eqnarray}
The matching of the inside and outside space-time demands the continuity of
the line element
\begin{equation} 
\label{stetig}
(\gamma \; {\mathrm{d}}u \; {\mathrm{d}}v)|_{r = R} = 
({\mathrm{d}}U \; {\mathrm{d}}V)|_{r = R} \;,
\end{equation}
which leads, with (\ref{ugenU}) and (\ref{ugenv}), to  
\begin{equation}
\label{z}
\frac{\mathrm{d}t}{\mathrm{d}\tau} = \frac{1}{\gamma(R)}
\sqrt{\gamma(R) (1 - \nu^2) + \nu^2}\; ,
\end{equation}
which describes the relation between $t$ and $\tau$, and finally fixes the 
space-time geometry. Note that Eq.~(\ref{z}) describes nothing but
the gravitational redshift $z$ for a photon that is emitted inside
the shell and crosses the shell at time $\tau$, i.e. 
$\mathrm{d}t/\mathrm{d}\tau = 1 + z$.

To trace a radial light ray through space-time, we need to find the transition 
functions between the two coordinate patches on the surface of the mass shell.
The setting is illustrated in Fig.~\ref{scene}. There are three different 
cases for a ray that arrives at infinity. Let us start with the trivial
case (3). Everything happens before the onset of expansion. We do not expect 
any effect then. For the second case (2) the rays enter the shell before 
the start of the expansion and the shell is `ignited' while the ray is crossing
the interior. Only a final intervall in $u$ of length $u_0$ is involved. 
For case (1), entrance and exit of the ray occur while the expansion is taking 
place.

We obtain a collapsing shell by inverting the setting 
($u\to-u, v\to-v, \tau \to -\tau$, etc.). To reproduce a collapse to a black 
hole, we need to take $R_0 \to 2M$ and $\nu \to 1$.\footnote{After this change 
of the time direction the in-vacuum is defined in a different region, but 
the vacuum expectation value of the number density operator is still given 
by the same formula.} This procedure will allow us to reproduce Hawking's
result from the setting of the expanding shell.

The transition functions, which follow from (\ref{ugenU}) and (\ref{ugenv}) by 
integration, are too complicated for an exact analytic treatment of the 
problem. We therefore have to find a reasonable approximation. Let us start 
from the inverse of Eqs.~(\ref{ugenU}) and (\ref{ugenv}), which reads
\begin{eqnarray}
\frac{{\mathrm{d}}U}{{\mathrm{d}}u} \bigg|_{U>-U_0}
             & = & \frac{[\gamma(1-\nu^2)+\nu^2]^\frac{1}{2}+\nu}{(1+\nu)}\ ,\\
\frac{{\mathrm{d}}V}{{\mathrm{d}}v} \bigg|_{V>0}
             & = & \frac{[\gamma(1-\nu^2)+\nu^2]^\frac{1}{2}-\nu}{(1-\nu)}\ .
\end{eqnarray}
As a first approximation we assume that the expansion velocity $\nu$ is 
constant. Secondly, we expect the biggest effect when the space-time 
curvature is high, thus in the vicinity of the `explosion'. Consequently we 
expand the transition functions in the Newtonian potential difference 
$ - 2M(1/R - 1/R_0)$, which is a small quantity at the beginning of 
the expansion. Case (3) is trivial. For case (2) we approximate 
$\mathrm{d}U/\mathrm{d}u$ around the `ignition', $R = R_0$. For case (1) we 
have to consider two points at which the ray crosses the shell. The effect 
from the entrance of the ray into the shell will be larger than the effect 
from the exit of the ray. Thus we need to approximate 
$\mathrm{d}V/\mathrm{d}v$ around $\gamma_0$ and $\mathrm{d}U/\mathrm{d}u$ 
around $\gamma(R(U=0)) =: \gamma_1$. Around $\gamma_0$ we include terms up to 
linear order, whereas around $\gamma_1$ we keep only the leading term. These 
expansions yield 
\begin{eqnarray}
\frac{{\mathrm{d}}U}{{\mathrm{d}}u} \bigg|_{U>-U_0} 
    &\approx& c_0 - c_0' \frac{2M}{R(U)}\; \; \mbox{around} \; \; \gamma_0\\ 
\frac{{\mathrm{d}}U}{{\mathrm{d}}u} \bigg|_{U>-U_0} 
    &\approx& c_1 \; \; \mbox{around} \; \; \gamma_1\\ 
\frac{{\mathrm{d}}V}{{\mathrm{d}}v} \bigg|_{V>0} 
    &\approx& d_0 - d_0' \frac{2M}{R(V)} 
                          \; \; \mbox{around} \; \; \gamma_0\;\;,
\end{eqnarray}
where
\begin{eqnarray} 
c_0  &:=& \frac{{\mathrm{d}}U}{{\mathrm{d}}u}\bigg|_{\gamma_0} +
          c_0'(1-\gamma_0)\ , \qquad 
c_0'  :=  \frac{\partial}{\partial\gamma}\,
          \frac{{\mathrm{d}}U}{{\mathrm{d}}u}\bigg|_{\gamma_0}\ ,\\
\label{c1}
c_1  &:=& \frac{{\mathrm{d}}U}{{\mathrm{d}}u}\bigg|_{\gamma_1}\ ,\\ 
\label{d}
d_0   &:=& \frac{{\mathrm{d}}V}{{\mathrm{d}}v}\bigg|_{\gamma_0} +
           d_0' (1-\gamma_0)\ , \qquad
d_0' := \frac{\partial}{\partial\gamma}\,
        \frac{{\mathrm{d}}V}{{\mathrm{d}}v}\bigg|_{\gamma_0}\ .
\end{eqnarray}
As a consequence
\begin{eqnarray}
\label{dudU}
\frac{{\mathrm{d}}u}{{\mathrm{d}}U} &\approx& 
\frac{R_0(1-\nu)+\nu(U+U_0)}{c_0[R_0(1-\nu)+\nu(U+U_0)]-c_0' 2M(1-\nu)} \\
\label{dvdV}
\frac{{\mathrm{d}}v}{{\mathrm{d}}V}&\approx& 
\frac{R_0(1+\nu)+\nu V}{d_0[R_0(1+\nu)+\nu V] - d_0' 2M(1+\nu)} \;\;.
\end{eqnarray}

We denote the approximate transition functions around $\gamma_0$ at the 
entrance with $\xi_0(V)$, around $\gamma_0$ at the exit with $\eta_0(u)$, 
and around $\gamma_1$ at the exit with $\eta_1(u)$. The inverse functions 
of the last two are written as $\eta_0^{-1}(U)$ and $\eta_1^{-1}(U)$, 
respectively.
Integration of (\ref{dudU}), (\ref{c1}) and  (\ref{dvdV}) provides the 
functions that will be relevant to our calculation
\begin{eqnarray} 
\label{uebergang}
\eta_0^{-1}(U) &=& \frac{U}{c_0} + \frac{c_0'2M (1-\nu)}{c_0^2\nu}
    \ln\left[1+\frac{c_0\nu U}{(c_0 R_0 - c_0'2M)(1-\nu) 
                               + c_0\nu U_0}\right],\\
\eta_1^{-1}(U) &=& \frac{U}{c_1},\\
\xi_0(V) &=& \frac{V}{d_0} +\frac{d_0'2M (1+\nu)}{d_0^2\nu}
    \ln\left[1+\frac{d_0\nu V}{(d_0 R_0 - d_0'2M)(1+\nu)}\right].
\end{eqnarray}

\section{Particle production}

To compute the particle production caused by the moving mass shell, we 
consider an observer who defines vacuum before the onset of expansion 
(at {\mcal I}$^-$) and ask how this vacuum is seen by an observer at
{\mcal I}$^+$, i.e.~we have to calculate the Bogoliubov transformations 
between the vacua at {\mcal I}$^-$ and {\mcal I}$^+$. We can do this 
by following the track of the light-rays. 

We restrict our attention to a massless, minimally coupled scalar field, 
which obeys the equation of motion $\Box \varphi = 0$. 
We expect a damping of the amplitude with $r^{-1}$ because of the spherical 
symmetry. Furthermore we split off an angular component in the form of 
spherical harmonics $Y_{lm}(\theta,\phi)$:
\begin{eqnarray}
\varphi(r,t,\theta,\phi)=
\sum_{l,m} \frac{1}{r}Y_{lm}(\theta,\phi)\Psi_l(t,r) \;\;.
\end{eqnarray}
With this ansatz, the wave equation in null coordinates becomes 
\begin{eqnarray} \label{wglSS}
\frac{\partial^2 \Psi_l}{\partial u \partial v}
=\gamma(r)\left( \frac{2M}{r^3} + \frac{l(l+1)}{r^2} \right)\Psi_l \;\;.
\end{eqnarray}
The right-hand side has a minimum outside the Schwarzschild radius and acts 
like a potential well. Because of this, incoming rays may be reflected at the 
gravitational potential. To a good approximation one may neglect this effect
of backscattering by simply dropping the right-hand side of (\ref{wglSS}).
In the same way we neglect potential terms in the inside. The index $l$ is left
out in the following, since $\Psi$ no longer depends on $l$ if the centrifugal 
potential is neglected.

\begin{figure}
\begin{center}
\unitlength1cm
\begin{picture}(13,4.5)
\put(0,3.5){$e^{-\mathrm{i}\omega v}$}
\put(1,3.7){\vector(1,0){1.2}}
\put(1.2,3.9){\scriptsize{$v>0$}}
\put(2.5,3.5){$e^{-\mathrm{i}\omega \xi(V)}$}
\put(4.2,3.7){\vector(1,0){1}}
\put(5.4,3.5){$e^{-\mathrm{i}\omega \xi(U)}$}
\put(7.2,3.7){\vector(1,0){1.2}}
\put(8.8,3.5){$e^{-\mathrm{i}\omega \xi(\eta(u))}$}
\put(11.2,3.5){case (1)} 
\put(1,3.5){\vector(1,-1){1.2}}
\put(0.6,2.6){\scriptsize{$v<0$}}
\put(2.5,2.0){$e^{-\mathrm{i}\omega V/\sqrt{\gamma_0}}$}
\put(4.2,2.2){\vector(1,0){1}}
\put(5.4,2.0){$e^{-\mathrm{i} \omega U/\sqrt{\gamma_0}}$}
\put(7.2,2.2){\vector(1,0){1.2}}
\put(7.3,2.5){\scriptsize{$U>-2R_0$}}
\put(8.8,2.0){$e^{-\mathrm{i}\omega \eta(u)/\sqrt{\gamma_0}}$}
\put(11.2,2.0){case (2)}
\put(7.2,2.0){\vector(1,-1){1.2}}
\put(6.3,1.0){\scriptsize{$U<-2R_0$}}
\put(7.5,0){$e^{-\mathrm{i}\omega (u + u_0 - 2R_0/\sqrt{\gamma_0})}$}
\put(11.2,0.2){case (3)}
\end{picture}
\end{center}
\caption{Tracing of the incoming ray.} 
\label{3faelle}
\end{figure}
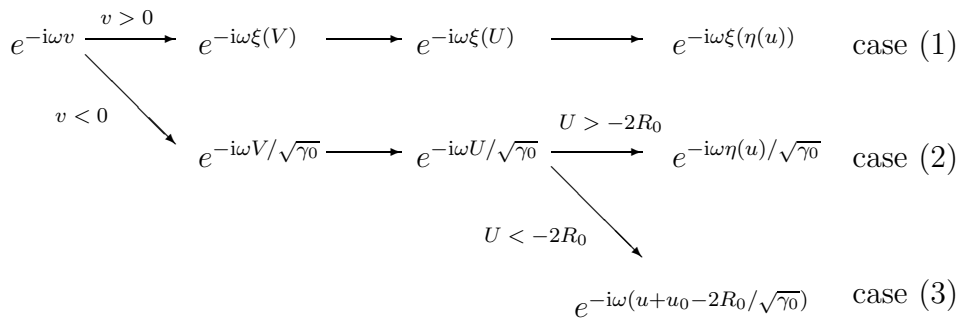

We are searching for solutions of the equations
\begin{eqnarray}
\frac{\partial^2 \Psi}{\partial u \partial v} =0
\; \; \;
\mbox{outside,}
\; \; \;
\frac{\partial^2 \Psi}{\partial U \partial V} =0
\; \; \;
\mbox{inside.}
\end{eqnarray}
Any function that does not mix the two variables $u,v$ or $U,V$ respectively
is a solution. We are looking for an incoming mode which turns into an 
outgoing one in a smooth manner and passes through $r=0$. These solutions 
must behave like spherically symmetric Minkowski modes at {\mcal I}$^-$. 
At $r \to \infty$ the phase of an incoming wave is given by
\begin{equation}
e^{- \mathrm{i}\omega v}\ .
\end{equation}
Inside of the mass shell, the corresponding mode travels on a curve with 
constant $V$. At $r=0$ we have $U=V$ where the incoming ray 
$\exp(- \mathrm{i} \tilde{\omega} V)$ turns smoothly into the outgoing
ray $\exp(- \mathrm{i} \tilde{\omega} U)$.
At {\mcal I}$^+$, outgoing modes are proportional to
\begin{equation}
e^{ -\mathrm{i}\omega' u}\ .
\end{equation} 
With the help of the transition functions (\ref{uebergang}) we can now 
glue curves of constant $v$ to curves of constant $V$ and curves of 
constant $U$ to curves of constant $u$. Figure \ref{3faelle} gives an 
overview of this procedure. Here we already inserted $U_0=2R_0$. 

In the following we restrict our analysis to the most interesting case (1), 
which provides the result for the extended expansion phase of the mass shell.  
Case (2) is restricted to a short time after the ignition of the shell, and 
thus is only characteristic of the ignition phase of our problem. Its 
importance depends on the velocity of the shell. So we now calculate the 
coefficients $\beta^1_{\omega'\omega}$. 
We expand the outgoing modes into plane waves. The outgoing modes are 
functions that depend on $u$ only, and we obtain 
\begin{equation}
\Psi \simeq \frac{1}{4\pi \sqrt{2\omega}} e^{-\mathrm{i}\omega\xi_0(\eta_1(u))}
= \int_{-\infty}^{\infty} A^1_{\omega}(\omega')\frac{1}{4 \pi \sqrt{2\omega'}}
e^{-{\mathrm{i}}\omega' u} \; {\mathrm{d}}\omega' \; \; \mbox{for} \; \; u>0\ ,
\end{equation}
with
\begin{eqnarray} 
\label{DefA1}
A^1_{\omega}(\omega') = \left\{ \begin{array}{rr}
\alpha_{\omega'\omega}^{1*} \; \; \omega' > 0 \\
{\mathrm{i}} \beta_{- \omega'\omega}^1 \; \; \omega' < 0
\end{array} \right. \;\;.
\end{eqnarray}
This is a Fourier transformation, which allows us to compute the Bogoliubov 
coefficients without evaluatiing the scalar product (\ref{skpr}). 
A Fourier transformation of the outgoing modes gives
\begin{equation}
I := \sqrt{\frac{\omega}{\omega'}}A^1_{\omega}(\omega')
     = \frac{1}{2\pi} \int_{0}^{\infty} 
        e^{-{\mathrm{i}}\omega \xi_0(\eta_1(u))} 
        e^{{\mathrm{i}}\omega' u} \; {\mathrm{d}}u \ .
\end{equation}
Using the approximations for $\xi_0(V)$ and $\eta_1(u)$, and introducing the 
notation $\lambda := d_0'2M(1+\nu)/(d_0^2\nu)$, we find
\begin{equation}
I = \frac{1}{2\pi} \int_{0}^{\infty}
      e^{-{\mathrm{i}}(\omega\frac{c_1}{d_0}-\omega')u} \left[ 
      \frac{(R_0 - d_0'2M/d_0)(1+\nu)/\nu + c_1 u}
           {(R_0 - d_0'2M/d_0)(1+\nu)/\nu}
      \right]^{-\mathrm{i}\omega\lambda} \; \mathrm{d}u \ .
\end{equation}
We further define $x-x_0 := c_1 u / d_0 $ and $x_0:=(R_0 -
d_0'2M/d_0)(1+\nu)/(\nu d_0)$.
The transition coefficients then become 
\begin{equation} 
\label{beta1}
\beta^1_{\omega'\omega} = 
-\frac{1}{2\pi}\sqrt{\frac{\omega'}{\omega}}
e^{\mathrm{i}\omega_+ x_0} x_0^{\mathrm{i}\omega\lambda} \frac{d_0}{c_1}
\int_{x_0}^{\infty} x^{-\mathrm{i}\omega\lambda} e^{-\mathrm{i}\omega_+ x} \; 
{\mathrm{d}}x \ ,
\end{equation}
where $\omega_+ := \omega + \omega' \frac{d_0}{c_1}$.

The vacuum expectation value of the number density operator is given by
\begin{equation} 
\label{Ablx}
\int_0^{\infty}\vert\beta^1_{\omega'\omega}\vert^2 \; {\mathrm{d}}\omega' =
\frac{d_0^2}{4\pi^2 c_1^2} \int_0^{\infty} \frac{\omega'}{\omega} 
\Biggl|\int_{x_0}^{\infty} e^{-\mathrm{i}\omega_+ x} 
x^{-\mathrm{i}\omega \lambda} \; {\mathrm{d}}x\Biggr|^2\;{\mathrm{d}}\omega'\ .
\end{equation}
This quantity is divergent and has to be regularized. First we will shift the 
$\omega'$-integration into the complex plane for some $\epsilon$. We do this 
because, for ${\rm Im}\, \omega_+ < 0$, the inner integrations are well defined 
and may be expressed as incomplete $\Gamma$-functions \cite{AS}. 
Further we regularize the number density with the help of a `frequency split'
\begin{eqnarray}
\vert \beta^1_{\omega'\omega} \vert^2 = \lim_{\omega \to \tilde{\omega}} 
\beta^1_{\omega'\omega} 
{\beta^1_{\tilde{\omega}\omega}}^*\ .
\end{eqnarray}
Finally, we take the difference to the limit $x_0 \to 0$, which allows us to
subtract the divergent part in a well defined manner. The point is that 
(\ref{Ablx}) reduces in this limit to the expression for the number density 
that shows up in the Hawking effect \cite{Haw},
\begin{equation}
\label{x0lim}
\lim_{x_0 \to 0}\int_0^{\infty} \vert\beta^1_{\omega'\omega}\vert^2 \;
{\mathrm{d}}\omega'  = \lim_{\tilde\omega \to \omega}
\frac{\delta(\tilde\omega - \omega)}{e^{2\pi\omega\lambda}-1}\ .
\end{equation}
The only difference is that $\lambda$ replaces the inverse of the surface 
gravity, $1/\kappa = 4 M$, that would show up in the situation of a collapsing
mass shell. 

As a first step we rotate the integration variable $x$ in the complex plane
and obtain 
\begin{equation}
\int_0^{\infty} \vert\beta^1_{\omega'\omega}\vert^2 \; {\mathrm{d}}\omega' = 
\lim_{\epsilon \to 0}
\frac{d_0^2}{4\pi^2 c_1^2} \int_{0-\mathrm{i}\epsilon}^{\infty-\mathrm{i}\epsilon} 
\frac{\omega'}{\omega} 
\Bigg|\int_{x_0}^{x_0-\mathrm{i}\infty} e^{-\mathrm{i}\omega_+ x} 
x^{-\mathrm{i}\omega\lambda} \; {\mathrm{d}}x \; \Bigg|^2
\;{\mathrm{d}}\omega' \; .
\end{equation}
Introducing $\tilde{\omega}_+ = \tilde{\omega} + 
\omega' \frac{d_0}{c_1}$ and $\tilde{\omega} := \omega + \varepsilon$,
we regularize the product of the inner integrations by a frequency split,
\begin{eqnarray} 
&&\int_0^{\infty} \vert \beta^1_{\omega'\omega} \vert^2 \; \mathrm{d}\omega' = 
\lim_{\epsilon \to 0 \atop \varepsilon \to 0}
\frac{1}{4\pi^2 c_1^2} \times \nonumber \\
&&\int_{0-\mathrm{i}\epsilon}^{\infty-\mathrm{i}\epsilon} 
\frac{\omega'}{\omega} 
\int_{x_0}^{x_0-\mathrm{i}\infty} e^{-{\mathrm{i}} \omega_+ x} 
x^{-\mathrm{i}\omega \lambda} \; {\mathrm{d}}x \; 
\int_{x_0}^{x_0+\mathrm{i}\infty} 
e^{\mathrm{i}\tilde{\omega}_+^* \tilde{x}} 
\tilde{x}^{\mathrm{i}\tilde{\omega} \lambda} \; \mathrm{d}\tilde{x} 
\;{\mathrm{d}}\omega' \; , \label{57}
\end{eqnarray}
where we took the limit partially. In the first of the inner integrals we 
further substitute $\mathrm{i}\omega_+ x =:y$ and
in the second $-\mathrm{i}\tilde{\omega}^*_+\tilde{x} =:\tilde{y}$, 
\begin{eqnarray} \label{55}
&&\int_0^{\infty} \vert \beta^1_{\omega'\omega} \vert^2 \; \mathrm{d}\omega' =
\lim_{\epsilon \to 0 \atop \varepsilon \to 0}
\frac{1}{4\pi^2 c_1^2 \omega} e^{- \pi \omega \lambda}\times \\
&&\int_{0-\mathrm{i}\epsilon}^{\infty-\mathrm{i}\epsilon} \omega' 
\frac{\omega_+^{\mathrm{i}\lambda\omega}}{\omega_+}
\frac{(\tilde{\omega}_+^*)^{-\mathrm{i}\lambda\tilde{\omega}}}
     {\tilde{\omega}_+^*}
\int_{\mathrm{i}\omega_+ x_0}^{\mathrm{i}\omega_+ x_0+\infty} e^{-y} 
y^{-\mathrm{i}\omega \lambda} \; {\mathrm{d}}y \; 
\int_{-\mathrm{i}\tilde{\omega}_+^* x_0}^{-\mathrm{i}\tilde{\omega}_+^* x_0+
\infty} e^{-\tilde{y}} 
\tilde{y}^{\mathrm{i}\tilde{\omega}\lambda} \; {\mathrm{d}}\tilde{y} \;
{\mathrm{d}}\omega' \; . \nonumber
\end{eqnarray}
Since
\begin{eqnarray}
(\omega_+^* + \varepsilon)^{-\mathrm{i}\tilde{\omega}\lambda}=
(\omega_+^*)^{-\mathrm{i}\tilde{\omega}\lambda }+O(\varepsilon) \;\;,
\end{eqnarray} 
and ignoring terms ${\cal O}(\varepsilon)$, we can write
\begin{eqnarray}
&&\int_0^{\infty} \vert\beta^1_{\omega'\omega}\vert^2 \; {\mathrm{d}}\omega' = 
\lim_{\epsilon \to 0 \atop \varepsilon \to 0}
\frac{d_0^2}{4\pi^2 c_1^2 \omega}e^{- \pi \omega \lambda}\times 
\nonumber \\
&&\int_{0-{\mathrm{i}}\epsilon}^{\infty-{\mathrm{i}}\epsilon} \omega'
\omega_+^{-\mathrm{i}\lambda\varepsilon-2}
\int_{\mathrm{i}\omega_+ x_0}^{\mathrm{i}\omega_+ x_0+\infty} e^{-y} 
y^{-\mathrm{i}\omega\lambda} \; {\mathrm{d}}y \; 
\int_{-{\mathrm{i}}\omega_+^* x_0}^{-\mathrm{i}\omega_+^* x_0+\infty} 
e^{-\tilde{y}} 
\tilde{y}^{{\mathrm{i}}\omega \lambda} \; 
{\mathrm{d}}\tilde{y} \;{\mathrm{d}}\omega' \; .
\label{beforesub}
\end{eqnarray}

Let us now take the difference to the $x_0 \to 0$ limit (\ref{x0lim}), 
\begin{eqnarray*}
&&\int_0^{\infty} \vert\beta^1_{\omega'\omega}\vert^2 \; \mathrm{d}\omega' -
\lim_{x_0 \to 0}\int_0^{\infty} \vert\beta^1_{\omega'\omega}\vert^2 \; 
{\mathrm{d}}\omega' =  
\lim_{\epsilon \to 0 \atop \varepsilon \to 0} 
\frac{d_0^2}{4\pi^2 c_1^2 \omega} e^{-\pi\omega\lambda}\times \nonumber \\
&&\int_{0-\mathrm{i}\epsilon}^{\infty-\mathrm{i}\epsilon} \omega' 
\omega_+^{-\mathrm{i}\lambda\varepsilon - 2} 
\left[\left|\int_{\mathrm{i}\omega_+ x_0}^{\mathrm{i}\omega_+ x_0+\infty} 
y^{-\mathrm{i}\omega\lambda} e^{- y} \mbox{d}y \right|^2\; - 
\vert \Gamma(1-\mathrm{i}\omega\lambda)\vert^2\right] \; \mbox{d}\omega' \; .
\end{eqnarray*}
Note that the expression inside the square bracket is well defined 
for all $\omega'$. We now replace the integration variable $\omega'$ by 
$\omega_+$, using its definition from above, and obtain (we omit 
$\epsilon$ and $\varepsilon$ in the discussion of the following step)  
\begin{eqnarray*}
&& 
\frac{1}{4\pi^2\omega} e^{-\pi\omega\lambda}
\left\{\int_{\omega}^{\infty} \omega_+^{- 1} 
\left[\left|\int_{\mathrm{i}\omega_+ x_0}^{\mathrm{i}\omega_+x_0+\infty} 
y^{-\mathrm{i}\omega\lambda} e^{-y} \mbox{d}y \; \right|^2 - 
\vert\Gamma(1-\mathrm{i}\omega\lambda) \vert^2 \right] \mbox{d}\omega_+ 
\right.\\ 
&& \left. - \omega \int_{\omega }^{\infty} \omega_+^{- 2} 
\left[\left|\int_{\mathrm{i}\omega_+ x_0}^{\mathrm{i}\omega_+ x_0 +\infty} 
y^{-\mathrm{i}\omega\lambda} e^{- y} \mbox{d}y \; \right|^2 
- \vert \Gamma(1-\mathrm{i}\omega\lambda)\vert^2 \right] 
\mbox{d}\omega_+ \right\} \ .
\end{eqnarray*}
The second term is finite (see Appendix) and, since we expect a divergent 
result, we can neglect it. The first term can be written as the integral 
from $0$ to $\infty$ minus the integral from $0$ to $\omega$. The latter 
integration is finite as well, since the difference of the $\Gamma$-functions 
vanishes as $\omega_+ \to 0$. We are left with
\begin{eqnarray*}
&&\int_0^{\infty} \vert \beta^1_{\omega'\omega} \vert^2 \; 
\mathrm{d}\omega' -
\lim_{x_0 \to 0}\int_0^{\infty} \vert\beta^1_{\omega'\omega}\vert^2 \;
\mathrm{d}\omega'
= 
\mbox{finite} +
\frac{1}{4\pi^2 \omega} e^{-\pi\omega\lambda}\times \nonumber \\
&& \int_0^{\infty} \omega_+^{- 1} 
\left[\left|\int_{\mathrm{i}\omega_+ x_0}^{\mathrm{i}\omega_+ x_0 + \infty} 
y^{-\mathrm{i}\omega\lambda} e^{- y} \mbox{d}y \; \right|^2 - 
\vert \Gamma(1-\mathrm{i}\omega\lambda) \vert^2 \right] \; 
\mbox{d}\omega_+ \ .
\end{eqnarray*}
Now we neglect all finite terms and use the fact that the second term inside
the square bracket is again (up to finite terms) the result of the 
$x_0 \to 0$ limit (\ref{x0lim}). We thus see that the divergent part of 
(\ref{beforesub}) can be written as  
\begin{eqnarray}
&&\int_0^{\infty} \vert \beta^1_{\omega'\omega} \vert^2 \; \mathrm{d}\omega'=
\lim_{\epsilon \to 0 \atop \varepsilon \to 0} 
\frac{1}{4\pi^2 \omega} e^{-\pi\omega\lambda}\times \nonumber \\
&&\int_{0-\mathrm{i}\epsilon}^{\infty-\mathrm{i}\epsilon}
\omega_+^{-\mathrm{i}\lambda\varepsilon-1}
\int_{\mathrm{i}\omega_+ x_0}^{\mathrm{i}\omega_+ x_0 + \infty} e^{-y} 
y^{-\mathrm{i}\omega \lambda} \; {\mathrm{d}}y \; 
\int_{-\mathrm{i}\omega_+^* x_0}^{-\mathrm{i}\omega_+^* x_0 + \infty} 
e^{-\tilde{y}} 
\tilde{y}^{-\mathrm{i} \omega \lambda} \; {\mathrm{d}}\tilde{y} \;
{\mathrm{d}}\omega_+ \ . 
\label{zurueck}
\end{eqnarray}
For positive $\varepsilon$ we can replace the contour parallel to the positive 
real axis by an integration along the negative imaginary axis. Then we 
substitute $e^z=\mathrm{i} \frac{\omega_+}{\omega}$ and have
\begin{eqnarray}
&&\int_0^{\infty} \vert \beta^1_{\omega'\omega} \vert^2 \; \mathrm{d}\omega' =
\lim_{\epsilon \to 0 \atop \varepsilon \to 0} 
\frac{1}{4\pi^2 \omega} e^{-\pi\omega\lambda}\times \nonumber \\
&&\int_{\ln \left(\frac{\epsilon}{\omega}\right)}^{\infty} e^{-\mathrm{i}\lambda \varepsilon z}
\int_{x_0 \omega e^z}^{x_0 \omega e^z+\infty} y^{- \mathrm{i}\omega\lambda} e^{- y} \mbox{d}y 
\; \int_{x_0 \omega e^{z*}}^{x_0 \omega e^{z*}+\infty} \tilde{y}^{\mathrm{i}\omega\lambda} 
e^{- \tilde{y}} 
\mbox{d}\tilde{y} \; \mbox{d} z \ .
\label{almostdone}
\end{eqnarray}
We can now take the limit $\epsilon \to 0$ and see that, since the integration 
is along the real axis, our integral can be written as
\begin{equation}
\int_0^{\infty} \vert \beta^1_{\omega'\omega} \vert^2 \; \mathrm{d}\omega' =
\lim_{\varepsilon \to 0} 
\int_{- \infty}^{\infty} e^{-\mathrm{i}\lambda \varepsilon z} h(z)\;
\mbox{d}z\ , \label{defh}
\end{equation}
with  
\begin{eqnarray}
h(z):=\frac{e^{-\pi\omega\lambda}}{4\pi^2 \omega} \int_{x_0 \omega e^z}^{\infty} 
y^{-\mathrm{i}\omega\lambda} e^{- y} \mbox{d}y \; 
\int_{x_0 \omega e^z}^{\infty} \tilde{y}^{\mathrm{i}\omega\lambda} e^{- \tilde{y}} 
\mbox{d}\tilde{y} \ ,
\end{eqnarray}
for any complex $z$. For real $z$, as we integrate over on the r.h.s.~of 
Eq.~(\ref{defh}), this yields the quantity we search for; $h(z)$ is identical 
to the product of the inner integrations in (\ref{almostdone}) for real $z$. 
This function is holomorphic in $z$. It has an upper bound $B$ for any real 
$z$ in $[-R,R]$. Thus  
\begin{eqnarray}
\left\vert \int_{-R}^R h(z) \mbox{d}z \right\vert \leq 2 R B \; \; ,
\end{eqnarray}
and, because the function is holomorphic, 
\begin{eqnarray}
\left\vert \int_{\gamma^u_{R}} h(z) \mbox{d}z \right\vert \leq 2 R B\quad 
\mbox{and} \quad 
\left\vert \int_{\gamma^d_{R}} h(z) \mbox{d}z \right\vert \leq 2 R B \;\; ,
\end{eqnarray}
where the path $\gamma^d_{R}$ is the half-circle of radius $R$ in the 
lower complex plane and $\gamma^u_{R}$ the half-circle of radius $R$ 
in the upper complex plane. From this we find
\begin{eqnarray}
\lim_{R \to \infty} \vert \int_{\gamma^u_{R}} \frac{h(z)}{z^2} \mbox{d}z \vert 
&=& 0 \\
\lim_{R \to \infty} \vert \int_{\gamma^d_{R}} \frac{h(z)}{z^2} \mbox{d}z \vert 
&=& 0 \ .
\end{eqnarray}
Our integral of interest can now be expressed as  
\begin{eqnarray}
\int_{-\infty}^{\infty}
e^{- \mathrm{i}\lambda \varepsilon z} h(z) \; \mbox{d} z = 
- \frac{1}{\lambda^2} \partial^2_{\varepsilon} \int_{-\infty}^{\infty}
e^{- \mathrm{i}\lambda \varepsilon z} 
\frac{h(z)}{z^2} \; \mbox{d} z  \; \; ,
\end{eqnarray}
which allows us to make use of the residue theorem.
Now for $\varepsilon > 0$ we may close the contour in the lower plane. 
To compute the integral we shift the pole into the lower plane. This yields
\begin{eqnarray}
\int_{-\infty}^{\infty}
e^{-\mathrm{i}\lambda \varepsilon z} \frac{h(z)}{z^2} 
\; \mbox{d} z = - 2\pi \left( \lambda \varepsilon h + \mathrm{i} \partial_z h
\right) \Theta(\varepsilon) \ .
\end{eqnarray}
With this we now get
\begin{eqnarray}
\int_0^{\infty} \vert \beta^1_{\omega'\omega} \vert^2 \; \mathrm{d}\omega'&=&
\lim_{\varepsilon \to 0} \frac{2\pi}{\lambda} h(0) \delta(\varepsilon) \\ 
&=& \lim_{\varepsilon \to 0} 
\frac{e^{-\pi\omega\lambda}}{2\pi\omega\lambda} 
\delta(\varepsilon) \int_{x_0 \omega }^{\infty} 
y^{-\mathrm{i}\omega\lambda} e^{- y} \mbox{d}y \; 
\int_{x_0 \omega }^{\infty} \tilde{y}^{\mathrm{i}\omega\lambda} e^{-\tilde{y}} 
\mbox{d}\tilde{y} \ , 
\end{eqnarray}
with the differentiation on the $\delta$-function interpreted as 
$0$.\footnote{It is an even function.} 
If we now put all pieces together, we finally find for the vacuum expectation 
value of the number density operator $N(\omega)$ [in case (1)]:
\begin{eqnarray} 
\int_0^{\infty} \vert \beta^1_{\omega'\omega} \vert^2 \; \mbox{d}\omega' &=&
\frac{e^{-\pi\omega \lambda}}{2\pi \lambda \omega } \delta(\varepsilon) 
\vert \Gamma(1-\mathrm{i}\omega\lambda,x_0 \omega)\vert^2
\; \; 
+ \mbox{finite} \nonumber \\ \label{erg1}
\Rightarrow \frac{N(\omega)}{{\bf{V}}}
&=&\frac{e^{-\pi\omega \lambda}}{2 \pi \lambda \omega } 
\vert \Gamma(1-\mathrm{i}\omega\lambda,x_0 \omega)\vert^2
\; \; .
\end{eqnarray}
The spectral energy flux through a 2-dimensional surface $\rho(\omega)$ then reads 
\begin{eqnarray} \label{energdi1}
\rho(\omega)&=& \frac{e^{-\pi\omega \lambda}\omega^2}{4\pi^3\lambda} 
\vert \Gamma(1+\mathrm{i}\omega\lambda,x_0 \omega)\vert^2\; .
\end{eqnarray}
 
We have made several assumptions, but some of them may be relaxed.  
First, we assumed that the velocity $\nu$ of the mass shell is constant. 
In a realistic scenario we would expect a high velocity at the beginning, 
decreasing towards a constant value (inertial motion), in contrast to a 
collapse where the velocity would increase. Secondly,
we approximated around the space-time points that cause the maximal amount 
of particle production. Therefore, our result is only correct in the vicinity
of $u=0$. To obtain a result for later times, we would have to approximate 
around other points. This gives rise to different values of the parameters 
$\lambda$ and $x_0$, which actually become functions of $R$ and $\nu(R)$. 
These functions then are [cf. (\ref{d})]:
\begin{eqnarray} \label{dR}
\lambda(R) &=& \frac{2 (1+\nu) M d'}{\nu d^2} 
\ \ \mbox{and} \nonumber \\
x(R) &:=& \frac{(1 + \nu) R}{\nu d} - \lambda \ ,\ \mbox{with} \nonumber \\
d'(R)&:=& \frac{\partial}{\partial
\gamma(R)} \frac{{\rm d} V}{ {\rm d} v} \bigg|_{\gamma(R)}   \nonumber\\
d(R) &:=&  \frac{{\rm d} V}{ {\rm d} v} \bigg|_{\gamma(R)} 
+ d'(R) [1-\gamma(R)]\ , 
\end{eqnarray}
from which we obtain a spectral energy flux that depends on the actual 
radius and the actual velocity of the mass shell:
\begin{eqnarray} \label{dasergebnis}
\rho(\omega,R) = \frac{e^{-\pi\omega\lambda(R)}\omega^2}
{4\pi^3\lambda(R)} 
\vert \Gamma[1+{\mathrm{i}}\omega\lambda(R), x(R)]\vert^2 \; \; .
\end{eqnarray}
The quality of this generalization mainly depends on the assumption 
that the change of velocity is small, so that $\ddot{R}$-contributions 
can be neglected. Should the acceleration of the mass shell get too high, 
the patches in which the velocity can be treated as nearly constant would 
get too small and effects from the boundaries would grow important.

\section{Discussion}

The particle production of an expanding mass shell depends on
the velocity of the mass shell $\nu$, its mass $M$, and its radius $R$.
In the spectral energy flux, these three quantities enter only through 
the functions $\lambda(\nu,M,R)$ and $x(\nu,M,R)$. To get a feeling for 
the behaviour of the function $\rho(\omega)$ we will first turn to a discussion
of these quantities. In Table 1 we provide the values of $\lambda$ and $x$ 
for some limits. The limit $\nu \to 1$ together with $R \to 2M$ reproduces
Hawking's result.  

\begin{table} \label{Limites}
\begin{center}
\begin{tabular}{|c|c|c|}
\hline
Limit &
$\lambda$ & $x$ \\
\hline \hline
$\nu \to 1$ & $4M$ & $2(R-2M)$ \\ \hline
$\nu \to 0 $ & $\infty$ & $\infty $ \\ \hline
$\nu \to 1 \wedge R \to 2M$ & $4M$ & $0$ \\
\hline
$R \to \infty$ & $\frac{(1+\nu)^2}{\nu}M$ & $\infty$ \\
\hline
$M \to 0$ & $0$ & $\frac{(1+\nu)}{\nu}R$ \\
\hline
$R \to 2M$ & $4M$ & $0$ \\
\hline
\end{tabular}
\end{center}
\caption{Some limiting cases for the functions that enter the expression for
the spectral energy flux through a 2-dimensional surface.}
\end{table}

\begin{figure} 
\centering
\epsfig{figure=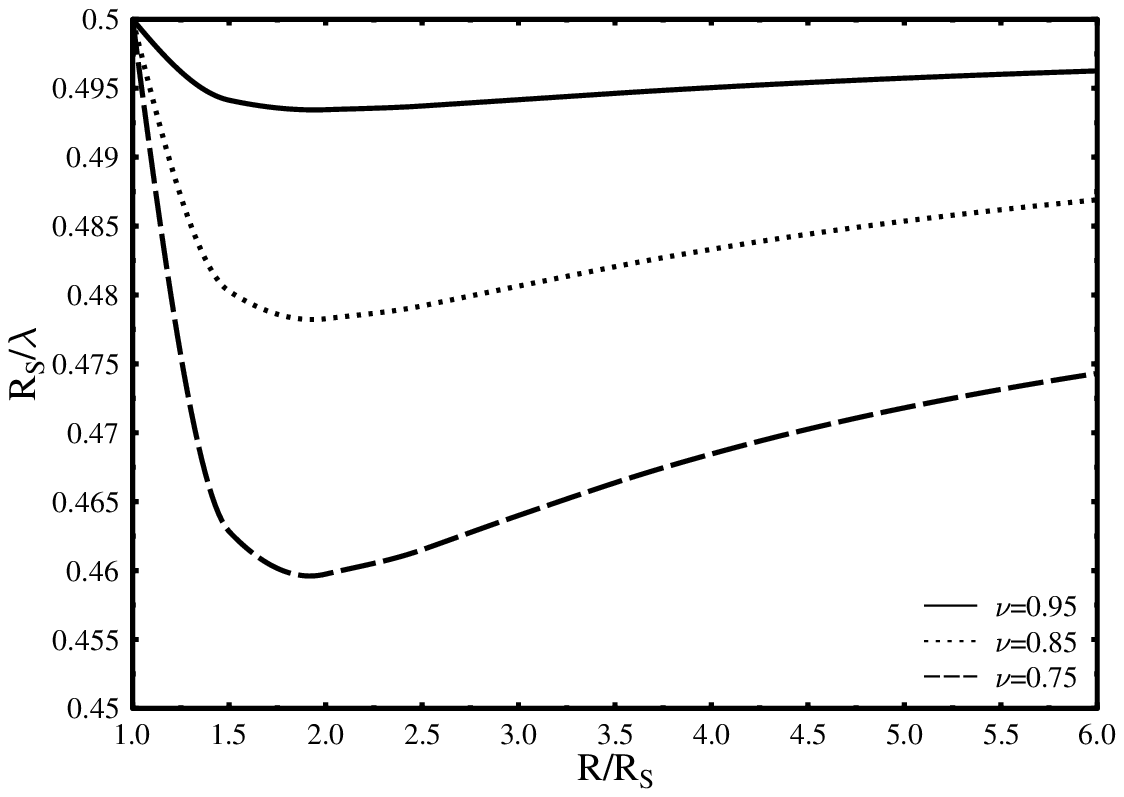,width=15cm}
\caption{The function $R_{\rm S}/\lambda(R)$ for different velocities $\nu$.
\label{la}}
 
\epsfig{figure=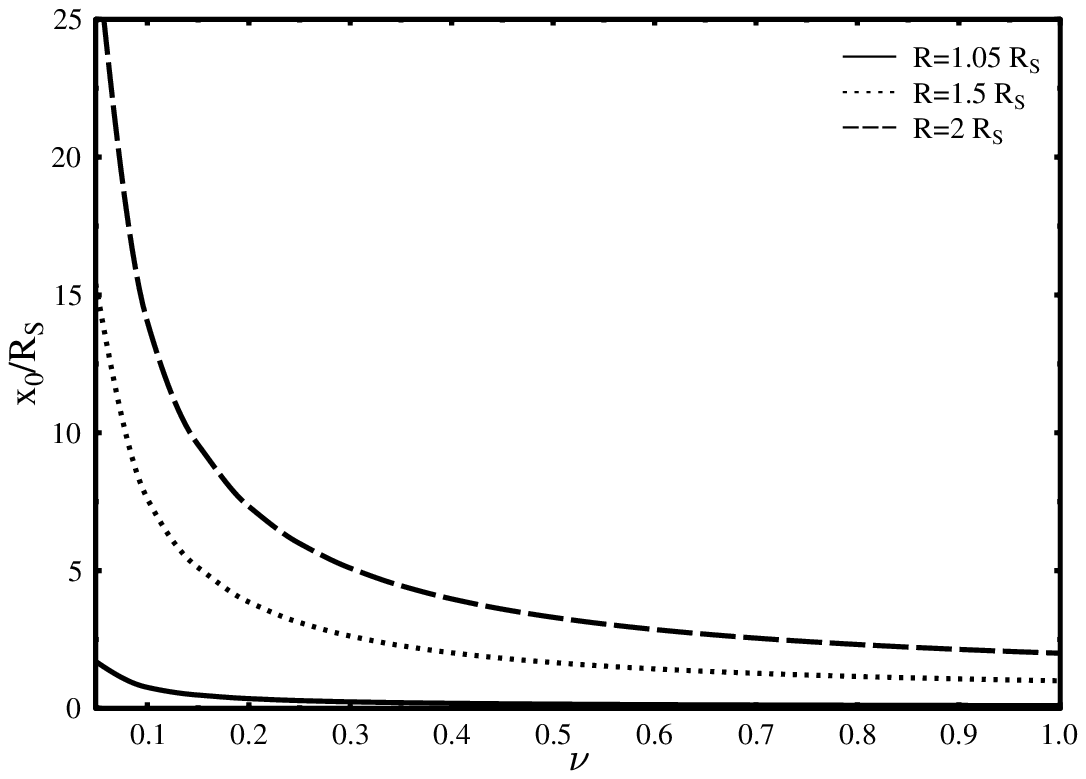,width=15cm}
\caption{The function $x(\nu)/R_{\rm S}$ for different values of the radius $R$.
\label{xnu}}
\end{figure}

In the Hawking limit we have $\lambda \to \kappa$, the so-called 
surface gravity. Additionally $x \to 0$, and we obtain exactly the 
expected Planck spectrum (cf. (\ref{dasergebnis})). 
Figures \ref{la} and \ref{xnu} show $1/\lambda$ as a function of $R$ for
three different values of $R$ and $x$ as a function of $\nu$ for different 
values of $R$. 
We normalize all quantities to the Schwarzschild radius $R_{\rm S} = 2M$;
$1/\lambda$ decreases rapidly with decreasing velocity. 
As is shown below, this implies a decrease of the spectral energy flux and a 
shift of the maximum towards smaller frequencies as the velocity decreases;
$\lambda$ is less sensitive to changes of the radius at high velocities. 
Particle production ceases, as expected, at large radii,   
because $x$ increases with large $R$ and small velocity.

\subsection{The spectral energy flux}

\begin{figure} 
\centering
\epsfig{figure=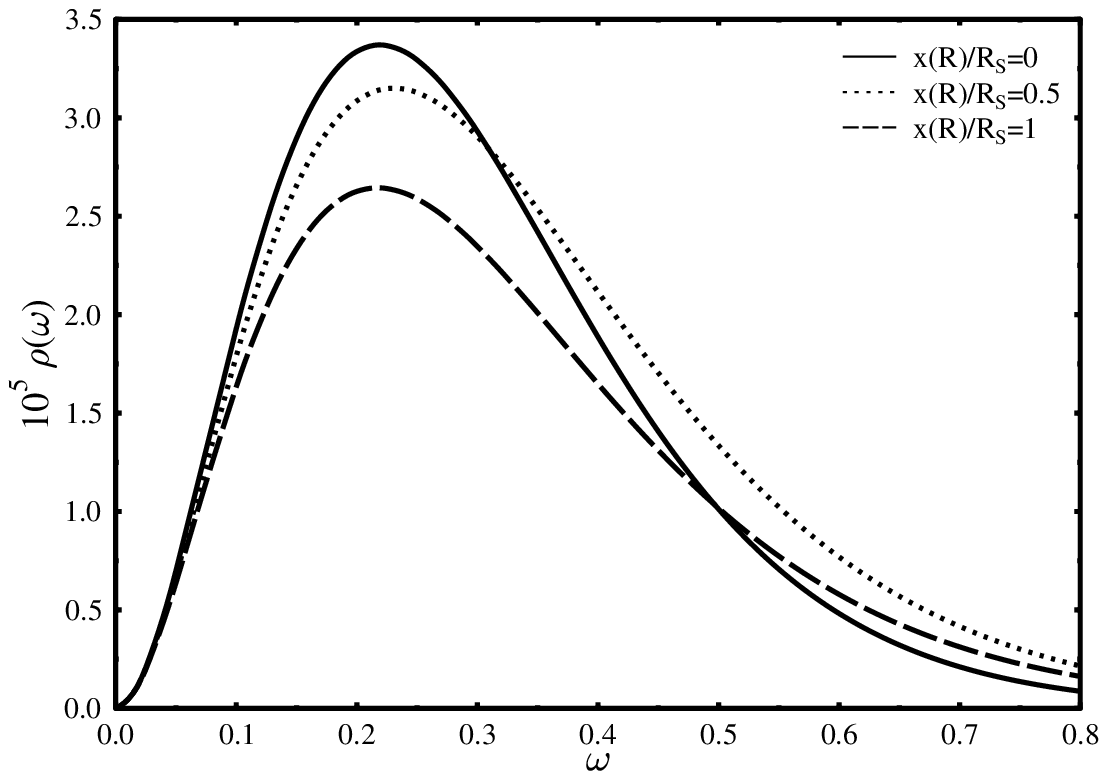,width=15cm} 
\caption{The spectral energy flux through a 2-dimensional surface $\rho(\omega)$ for different values 
of $x/R_{\rm S}$ at fixed $\lambda/R_{\rm S}=2.05$.}
\label{Erg1}
 
\epsfig{figure=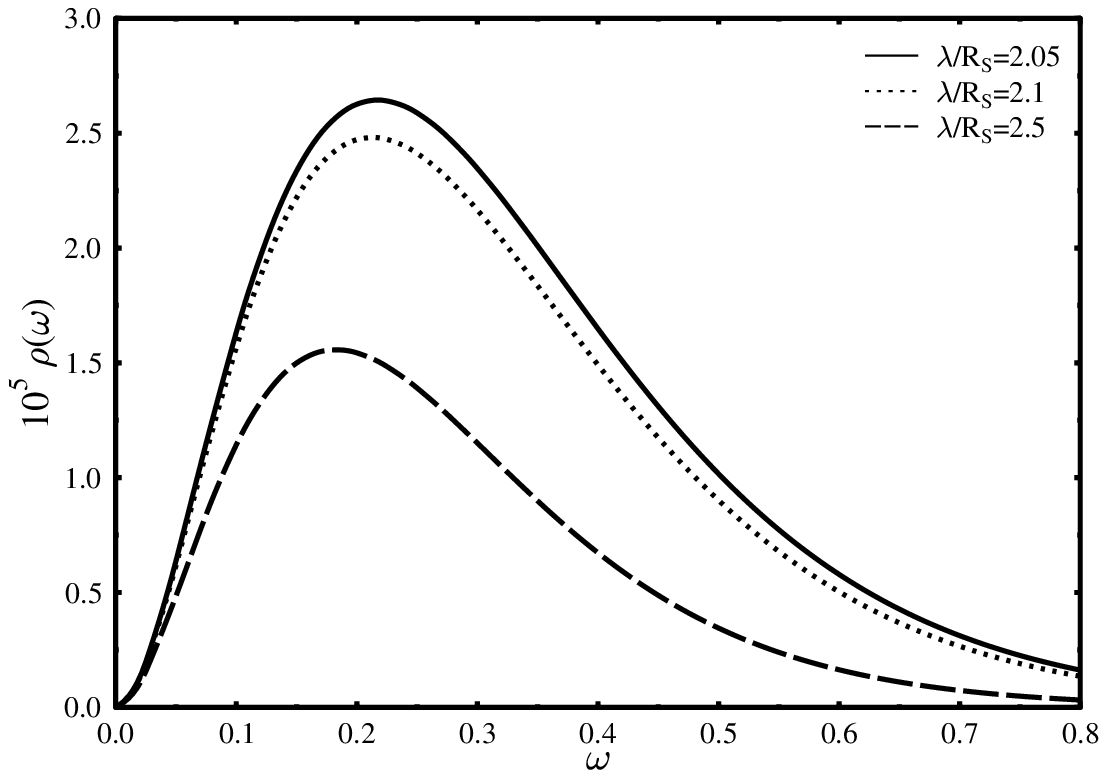,width=15cm}
\caption{The spectral energy flux through a 2-dimensional surface $\rho(\omega)$ for different values 
of $\lambda/R_{\rm S}$ at fixed $x/R_{\rm S} =1$.}
\label{Erg2}
\end{figure}

We now study the behaviour of the spectral energy flux (\ref{energdi1}) as 
a function of $\lambda/R_{\rm S}$ and $x/R_{\rm S}$. Figures \ref{Erg1} and 
\ref{Erg2} show $\rho(\omega)$ for different values of $x/R_{\rm S}$, and for 
different values of $\lambda/R_{\rm S}$ with the second parameter fixed at a 
representative value each time.
Let us first turn to Fig.~\ref{Erg1}. 
When $x=0$ we find a Planck spectrum with temperature 
$T=(2 \pi k_B \lambda )^{-1}$. For $x/R_{\rm S}>0$ the spectrum is non-thermal.
It is interesting to note that in the non-equilibrium situation more hard 
particles and less soft particles are produced at a fixed value of $\lambda$. 
 
In Fig.~\ref{Erg2} we see, as expected, that an increase in $\lambda/R_{\rm S}$ 
causes a shift of the maximum towards smaller frequencies. 
Furthermore the whole curve flattens quickly for increasing $\lambda/R_{\rm S}$.

\begin{figure} 
\centering
\epsfig{figure=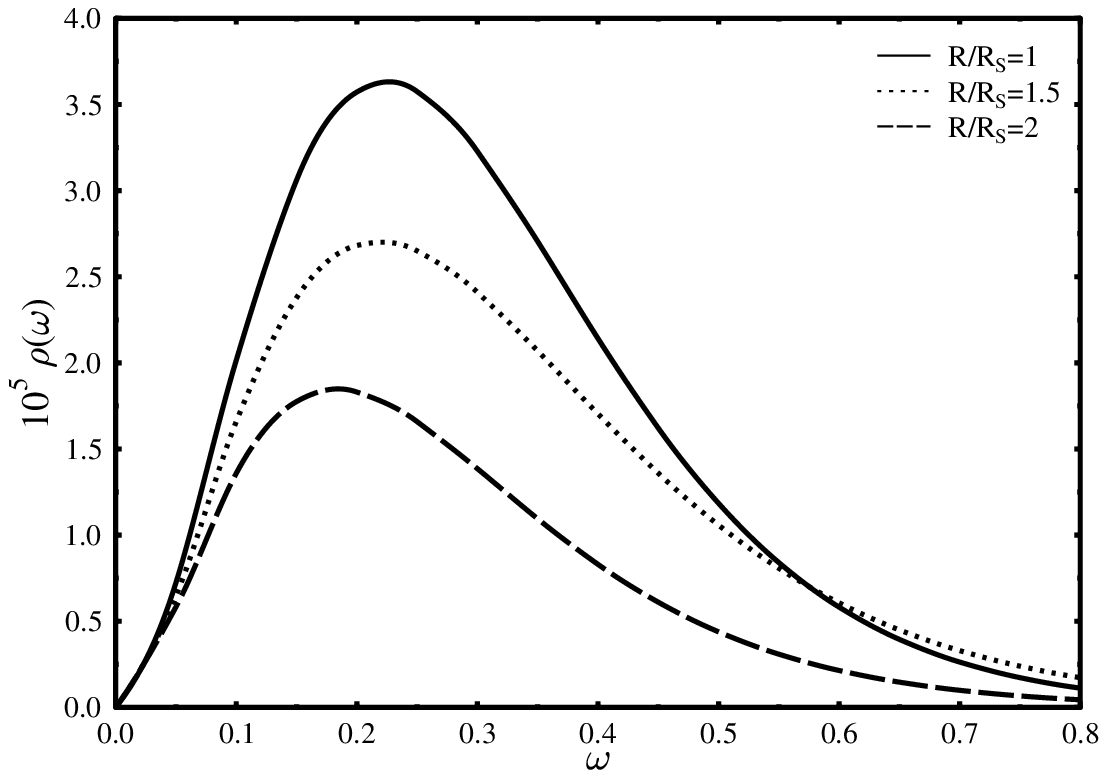,width=15cm}
\caption{The spectral energy flux through a 2-dimensional surface $\rho(\omega)$ at different moments
of time, respresented by three positions $R/R_{\rm S}$, for the constant 
velocity 
$\nu=0.95$.
\label{Erg4}}

\epsfig{figure=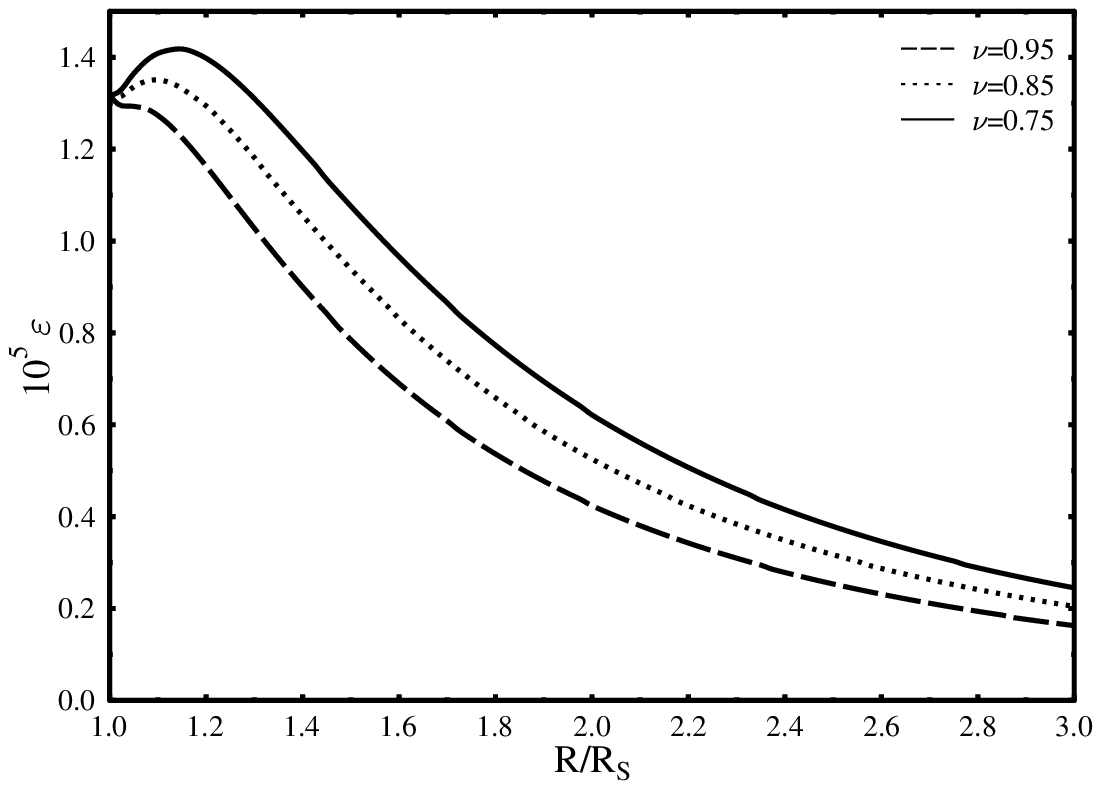,width=15cm}
\caption{The energy flux through a 2-dimensional surface for different values of the velocity $\nu$.
\label{Gesamten}}
\end{figure}

Remember that, according to the considerations presented in the last section, 
an increasing radius (which changes $\lambda$ and $x$) 
dominates the evolution of the spectral energy flux during the expansion. 
This can be seen in Fig.~\ref{Erg4}.

\subsection{The energy flux}

Finally we come to the emitted energy flux through a 2-dimensional surface, $\varepsilon \equiv 
\int\rho(\omega){\rm d}\omega$.

In Fig.~\ref{Gesamten} we see the energy flux as a function of radius 
for different values of $\nu$. At large values of $R/R_S$ no particles are
produced, since the mass shell moves inertially (the tidal forces go to zero). 
However, more interesting is the maximum of the curve that appears for 
ultra-relativistic velocities $\nu$. This reflects the fact that two effects 
are working against each other during the expansion. The probability of 
particle production depends on the strength of the tidal forces acting
on the vacuum locally. The tidal forces decrease monotonically  
as the mass shell expands. Once particles have been created, they have to 
climb out of the gravitational potential well of the mass shell itself
(gravitational redshift). This effect reduces the energy of the produced 
particles. At the Schwarzschild radius the energy would be reduced to zero.
For an ultrarelativistic mass shell the tidal forces change very quickly and 
thus the particle production is large. Close to the Schwarzschild radius 
however, this is counteracted by the gravitational redshift,
giving rise to a maximum slightly away from $R_{\rm S}$.  

\section{Summary}

We computed the spectrum of particles produced by the time-dependent 
gravitational field of an expanding mass shell. 
This system is---as long as no horizon forms---invariant under time-reversal, 
so that we can describe an expanding shell as well as a collapsing shell before 
the formation of a horizon. If the horizon is formed, the system gets 
quasi-static for the asymptotic observer and the effect is ``frozen''. In this
way we can reproduce the result of Hawking radiation. 

In the situation of an expanding mass shell the spectrum is non-thermal.
Our result might be of relevance for the discussion of evaporating black
holes. In the situation of a collapsing mass shell, we obtain the 
deviations from the Planck spectrum of Hawking radiation, which is 
produced just before the horizon forms. 

\section*{Acknowledgements}

DJS thanks the Austrian Academy of Sciences for financial support.

\begin{appendix}
\section{}

To show that 
\begin{eqnarray} \label{beid}
\int_\omega^{\infty}\!\!\! \omega_+^{- 2}\! 
\left[\int_{{\mathrm i}x_0 \omega_+}^{\mathrm{i}\omega_+x_0 + \infty}\!\!\!\!\! 
\! y^{\mathrm{i}\omega\lambda} e^{-y} \mbox{d}y  
\int_{- {\mathrm i}x_0\omega_+}^{-\mathrm{i}\omega_+x_0 + \infty}\!\!\!\!\!\!
\tilde{y}^{-\mathrm{i}\omega\lambda} 
e^{-\tilde{y}}\mbox{d}\tilde{y}
- \vert \Gamma(1+\mathrm{i}\omega\lambda)\vert^2 \right] 
\mbox{d}\omega_+
\end{eqnarray}
is finite (for $\omega>0$), it is sufficient to show that
\begin{eqnarray} \label{zeig}
\left\vert 
\int_{{\mathrm i}x_0\omega_+}^{\mathrm{i}\omega_+x_0 + \infty} 
y^{\mathrm{i}\omega\lambda} e^{-y} \mbox{d}y \right\vert 
\leq {\mbox{const.}},
\end{eqnarray}
which means that it has an upper bound for $\omega_+ \to \infty$ since the 
second term in (\ref{beid}) does not depend on $\omega_+$ anyhow. We are 
then left with an integral over $\omega_+^{-2}$ that is finite. If we succeed 
in showing that (\ref{zeig}) is true, we know that the first term does not 
increase faster for large $\omega_+$ and therefore is finite too.
It is
\begin{eqnarray}
\Bigg| 
\int_{{\mathrm i}x_0\omega_+}^{\mathrm{i}\omega_+x_0 + \infty} y^{\mathrm{i}\omega\lambda} 
e^{-y} \mbox{d}y \Bigg| &=&
\Bigg| 
\int_{{\mathrm i} x_0\omega_+}^{\mathrm{i}\omega_+x_0 + \infty} 
e^{\mathrm{i}\omega\lambda\ln\vert y \vert -\omega\lambda{\mathrm{arg}}(y)} 
e^{-y} \mbox{d}y \Bigg|\\
&\leq&\sup_{y \in \gamma}\Bigg|
e^{\mathrm{i}\omega\lambda\ln\vert y \vert -
\omega\lambda\mathrm{arg}(y)}\Bigg|\cdot
\Bigg| 
\int_{x_0\omega_+}^{-\mathrm{i}\omega_+\infty} 
e^{- y} \mbox{d}y \Bigg|\;\;,
\end{eqnarray}
$\gamma$ being the path of integration in the upper right 
quadrant of the complex plane and therefore is $\mathrm{arg}(y) \in 
\left[\frac{3}{2}\pi,2\pi\right]$ on $\gamma$. This yields
\begin{eqnarray}
\Bigg| 
\int_{{\mathrm i}x_0\omega_+}^{\mathrm{i}\omega_+x_0 + \infty} y^{\mathrm{i}\omega\lambda} 
e^{- y} \mbox{d}y \Bigg|
&\leq& 
e^{2 \pi \omega\lambda}
\Bigg| 
\int_{{\mathrm i}x_0\omega_+}^{\mathrm{i}\omega_+x_0 + \infty} 
e^{- y} \mbox{d}y \Bigg|\\
&=&
e^{2\pi\omega\lambda}
\Bigg| e^{-{\mathrm i}x_0\omega_+}\Bigg| \; \; .
\end{eqnarray}
Putting this together we have 
\begin{eqnarray}
\Bigg| 
\int_{\mathrm{i}x_0\omega_+}^{-\mathrm{i}\omega_+\infty} 
y^{\mathrm{i}\omega\lambda} 
e^{- y} \mbox{d}y \Bigg| \leq 
e^{2\pi\omega\lambda} \ ,
\end{eqnarray}
which shows that (\ref{beid}) is finite, as claimed.

\end{appendix}

\end{document}